\documentclass[12pt,onecolumn]{article}
\makeatletter
\let\frontmatter@title@above=\relax
\makeatother
\usepackage{amsmath}	% Advanced maths commands
\usepackage{appendix}
\usepackage[english]{babel}
\usepackage{blindtext}
\usepackage{color}
\usepackage{graphicx}	% Including figure files
\usepackage{textgreek}
\usepackage[normalem]{ulem}
\usepackage{xcolor}
\usepackage{tikz}
\usepackage{subcaption}
\usepackage[letterpaper,margin=1in]{geometry}
\usepackage{url}
\usepackage{hyperref}

\makeatletter
\let\@fnsymbol\@arabic
\makeatother

\title{Image Triangulation Using the Sobel Operator for Vertex Selection}

\author{Olivia X. Laske\thanks{Department of Mathematics, Statistics, and Computer Science, Macalester College, 1600 Grand Ave, Saint Paul, MN 55105, USA} \and Lori Ziegelmeier$^1$}

\date{}

\begin{document}
\maketitle

\begin{abstract}
    Image triangulation, the practice of decomposing images into triangles, deliberately employs simplification to create an abstracted representation. While triangulating an image is a relatively simple process, difficulties arise when determining which vertices produce recognizable and visually pleasing output images. With the goal of producing art, we discuss an image triangulation algorithm in Python that utilizes Sobel edge detection and point cloud sparsification to determine final vertices for a triangulation, resulting in the creation of artistic triangulated compositions.
\end{abstract}

\section{Introduction}
\label{sec:typesetting-summary}
Image triangulation creates an abstract representation of an image, which can be defined by four primary principles: the output image (a) divides the original image into a set of non-overlapping triangles, (b) simplifies the original image, (c) approximates original image features as triangles, and (d) retains the integrity of the original image. Several algorithms exist to perform image triangulation, such as \cite{Lawonn}, \cite{Marwood}, \cite{Onoja}, and \cite{Simo}. While this may be done for a variety of purposes, our goal is to use image triangulation as a form of art.

Here, we implement an image triangulation algorithm to create visual works of art; see GitHub repository \cite{github}. To simplify processing, the algorithm converts an image to grayscale. Next, it sharpens the image and applies the Sobel operator to detect edges. After identifying key pixels along the edges, it triangulates them and fills the resulting triangles with color to create the triangulated image. We discuss each step of our approach in the following section.

\section{Image Triangulation Algorithm}
\begin{figure}[thbp]
\begin{minipage}[t]{.49\linewidth}
\centering\includegraphics[width=\textwidth]{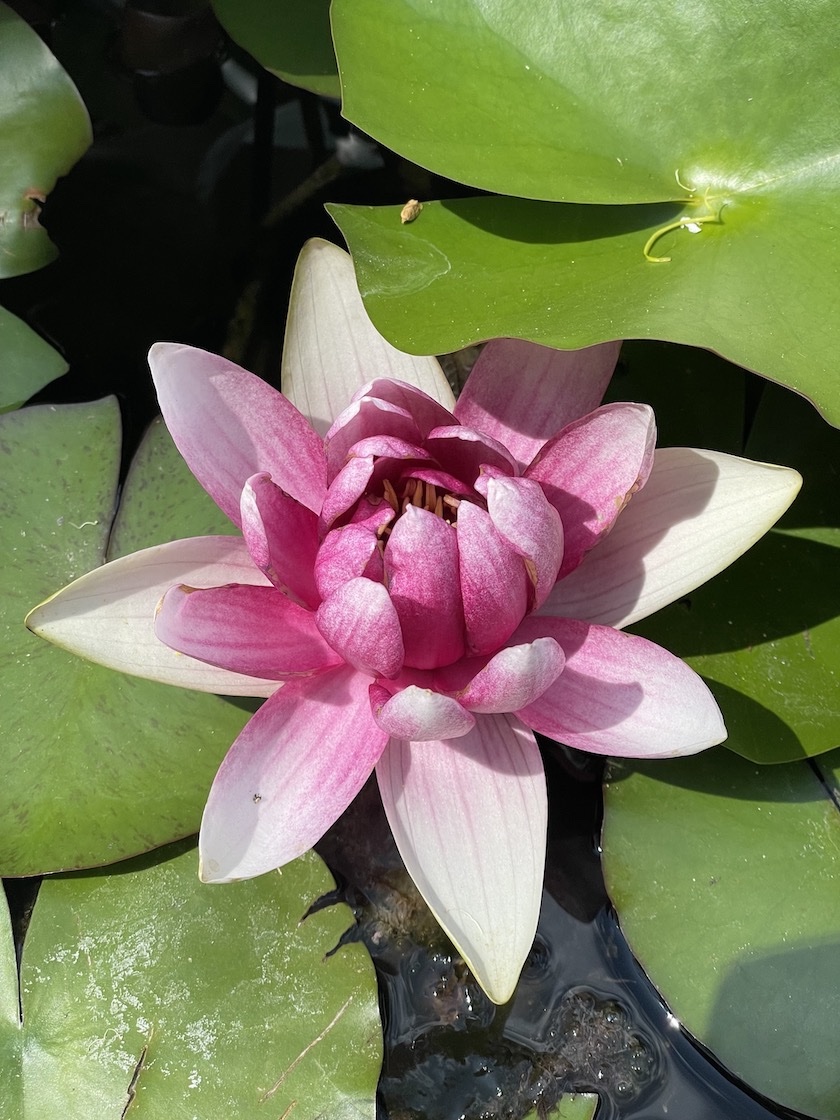}
%\subcaption{a short caption}\label{fig:2:1}
\end{minipage}
\begin{minipage}[t]{.49\linewidth}
\centering\includegraphics[width=\textwidth]{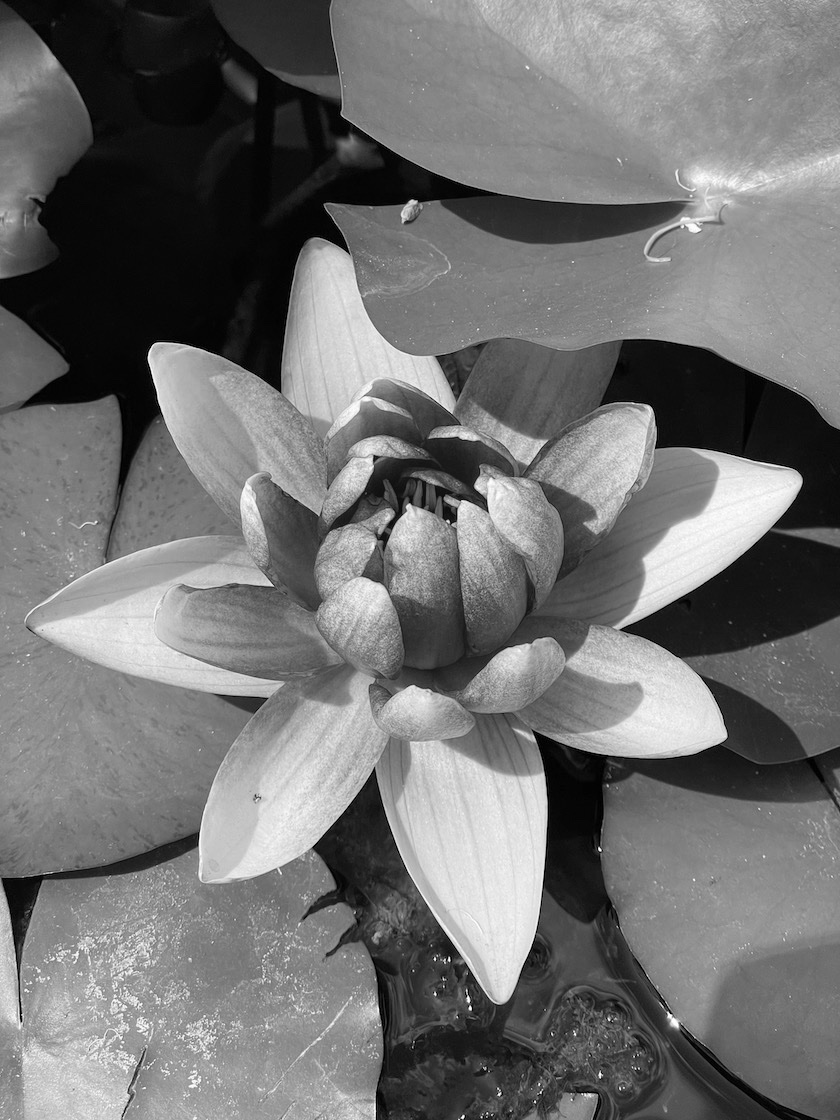}
%\subcaption{}\label{fig:2:2}
\end{minipage}
\caption{(Left) Original image before processing. (Right) Image after converting to grayscale.}
\label{fig:ImageandGray}
\end{figure}

\subsection{Convert to Grayscale}
We illustrate our algorithm with an image taken by the first author, Figure \ref{fig:ImageandGray} (Left). We convert the original image to grayscale using the linear combination for the red, green, and blue (RGB) pixel values: $c=0.299R+0.587G+0.114B$, where $c$ is rounded to the nearest integer \cite{Pham}. After iterating over each pixel, Figure \ref{fig:ImageandGray} (Right) displays the grayscale image.

\subsection{Sharpen Image} 
\begin{figure}[thbp]
\begin{minipage}[t]{.49\linewidth}
\centering\includegraphics[width=\textwidth]{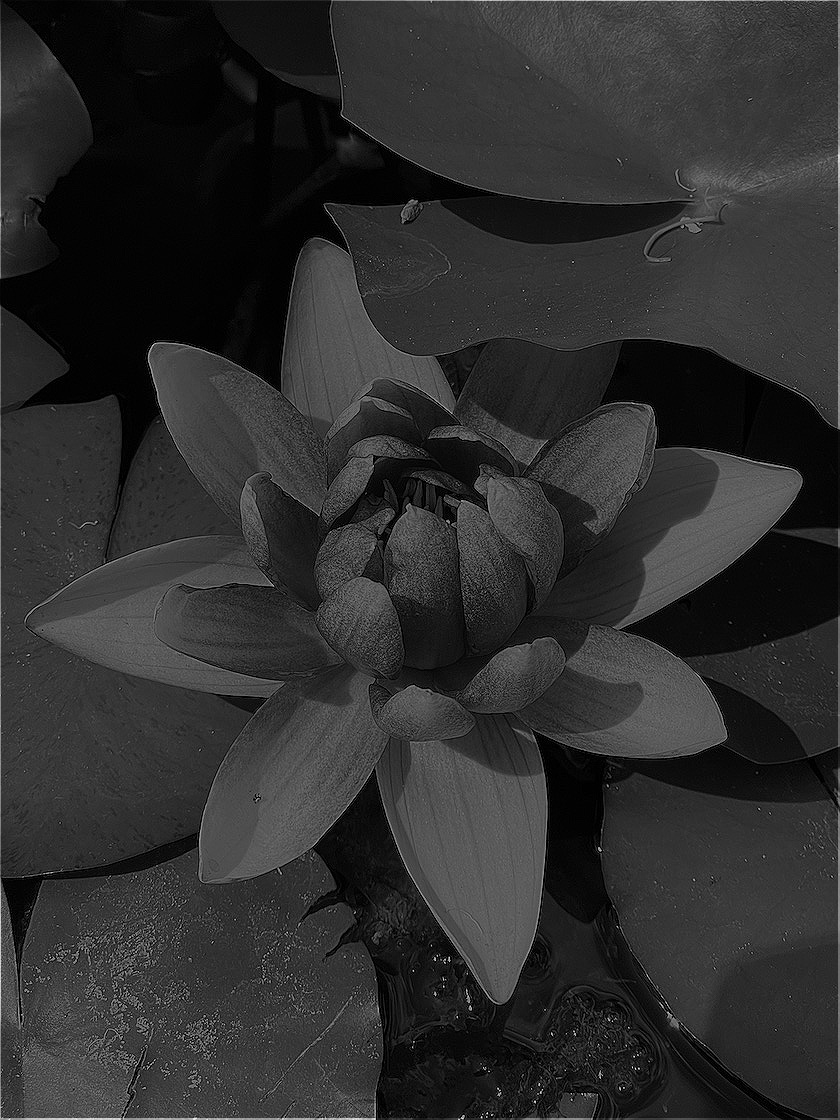}
%\subcaption{a short caption}\label{fig:2:1}
\end{minipage}
\begin{minipage}[t]{.49\linewidth}
\centering\includegraphics[width=\textwidth]{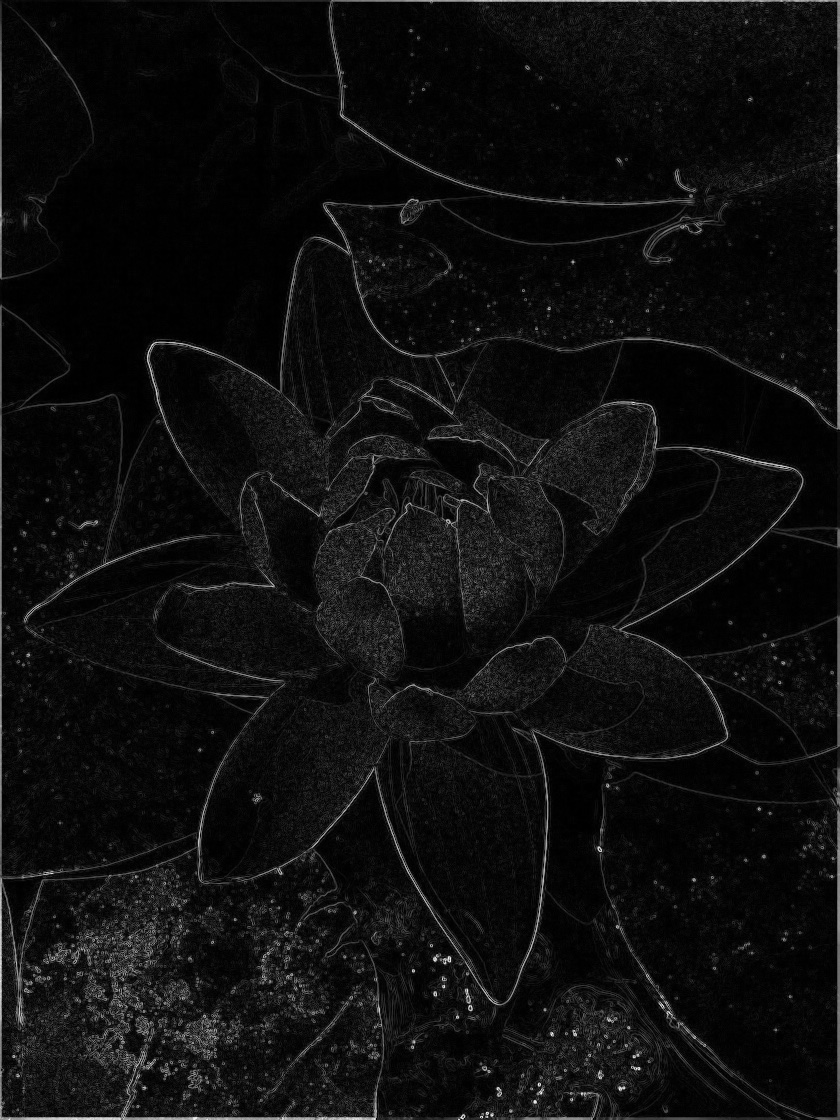}
%\subcaption{}\label{fig:2:2}
\end{minipage}
\caption{Image after (Left) sharpening with Eq. \ref{eq:sharpeningKernel} kernel, (Right) applying the Sobel operator.}
\label{fig:SharpenSobel}
\end{figure}

Next, we use a Laplacian sharpening kernel $L$ to more clearly define the edges \cite{Pham}:
\begin{equation}\label{eq:sharpeningKernel}
    L=\begin{bmatrix} 0 & -1 & 0 \\ -1 & 5 & -1 \\ 0 & -1 & 0 \end{bmatrix}.
\end{equation}
\begin{figure}
    \centering
    \includegraphics[width=\textwidth]{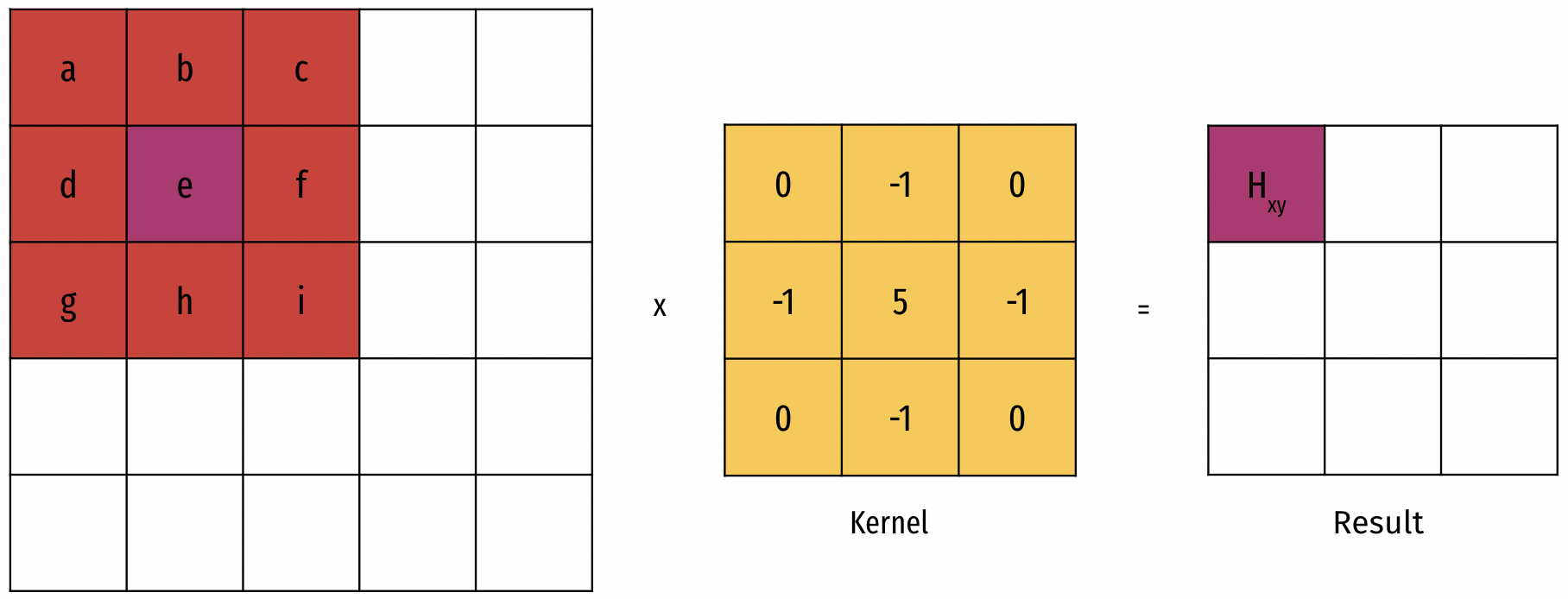}
    \caption{The image convolution process, which produces a new image by taking a weighted sum of each original pixel and its surrounding pixels. For the pink pixel $I_{xy}$, $H_{xy}=-b-d+5e-f-h$.}
    \label{fig:convolution}
\end{figure}
We use $L$ to perform image convolution, demonstrated in Figure \ref{fig:convolution}. For each interior pixel $I_{xy}$ of the $n\times n$ image $I$, we calculate the sharpened grayscale value $H_{xy}$ as the weighted sum of $I_{xy}$ and the surrounding pixels (assuming 0-indexing), 
\begin{equation}\label{eq:sharpening}
    H_{xy}=\sum_{i=-1}^1\sum_{j=-1}^1I_{x+i,y+i}L_{i+1,j+1}.
\end{equation}
Repeating the process for each interior pixel, we create a new sharpened image of size $(n-2)\times(n-2)$ since boundary pixels are not mapped to the new image. Figure \ref{fig:SharpenSobel} (Left) displays the image after using the sharpening procedure.

\subsection{Apply Sobel Operator}
We then apply the Sobel operator to extract the edges in the image \cite{Tian}. The process requires running the image convolution process twice using the following $x$ and $y$ kernels,
\begin{gather}\label{eq:sobel}
    G_x=\begin{bmatrix} 1 & 0 & -1 \\ 2 & 0 & -2 \\ 1 & 0 & -1 \end{bmatrix}, \quad 
    G_y=\begin{bmatrix} 1 & 2 & 1 \\ 0 & 0 & 0 \\ -1 & -2 & -1 \end{bmatrix}, \quad
    G=\sqrt{G_x^2+G_y^2}.
\end{gather}
The x-kernel whitens pixels to the left and right of pixel $I_{xy}$, while the y-kernel whitens pixels above and below pixel $I_{xy}$. Pixels that are part of a well-defined edge in the resulting $(n-4) \times (n-4)$ image\footnote{Alternatively, one could adjust the indexing of both the sharpening kernel and Sobel operator to include boundary pixels.} have grayscale values closer to white (255) than to black (0); see Figure \ref{fig:SharpenSobel} (Right). We reject pixels below a threshold grayscale value $t$, selecting $t=50$.

\subsection{Triangulate Points}
\begin{figure}[thbp]
\begin{minipage}[t]{.49\linewidth}
\centering\includegraphics[width=\textwidth]{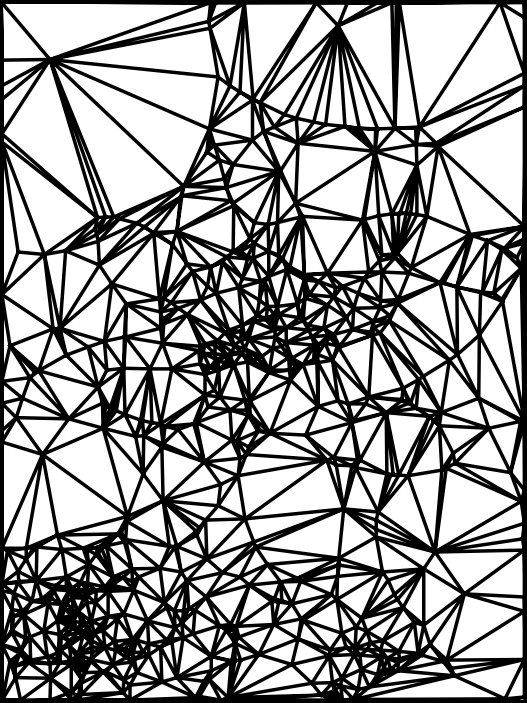}
%\subcaption{a short caption}\label{fig:2:1}
\end{minipage}
\begin{minipage}[t]{.49\linewidth}
\centering\includegraphics[width=\textwidth]{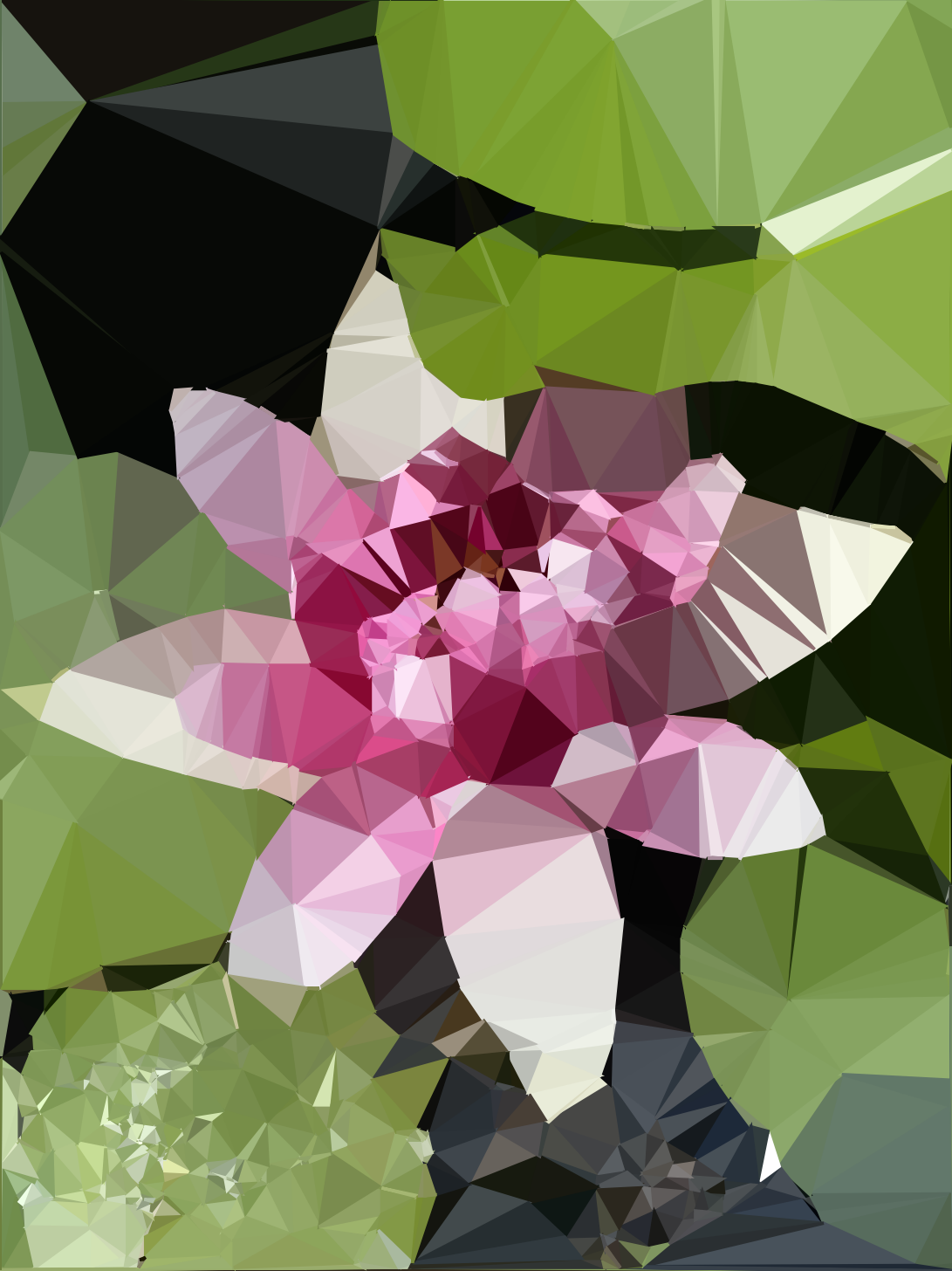}
%\subcaption{}\label{fig:2:2}
\end{minipage}
\caption{(Left) Triangulation of image. (Right) Final colored image triangulation.}
\label{fig:TriangulationColor}
\end{figure}

Let $S$ be the set of pixels above the Sobel threshold; these are used as the vertices to create the triangulation. Even with a threshold value, the pixels are too densely packed to create a visually appealing image. We reduce the density by uniformly sampling $|S|/d$ points, where $|S|$ is the number of pixels in $S$ and $d$ is a density parameter specified by the user. We select $d=60$. From these sampled points as vertices, we then compute the Delaunay triangulation. This choice can be modified by the user if there is a preference for a particular triangulation behavior, such as anisotropy. Figure \ref{fig:TriangulationColor} (Left) shows this triangulation.

\subsection{Color in Triangles}
Finally, we calculate the centroid of each triangle with the average of its vertex coordinates, rounded to the nearest integer, to obtain indices $(x,y)$. The final color of the triangle is equal to the RGB value at pixel $(x,y)$ in the original image. Figure \ref{fig:TriangulationColor} (Right) depicts the final image triangulation.

\section{Varying Triangulation Parameters}
We investigate the effect of varying two parameters selected by the user: the threshold value $t$ to select edges after applying the Sobel operation and the density parameter $d$ to subsample pixels above this threshold.
\begin{figure}[thbp]
    \begin{subfigure}[t]{0.3\textwidth}
        \centering
        \includegraphics[width=\textwidth]{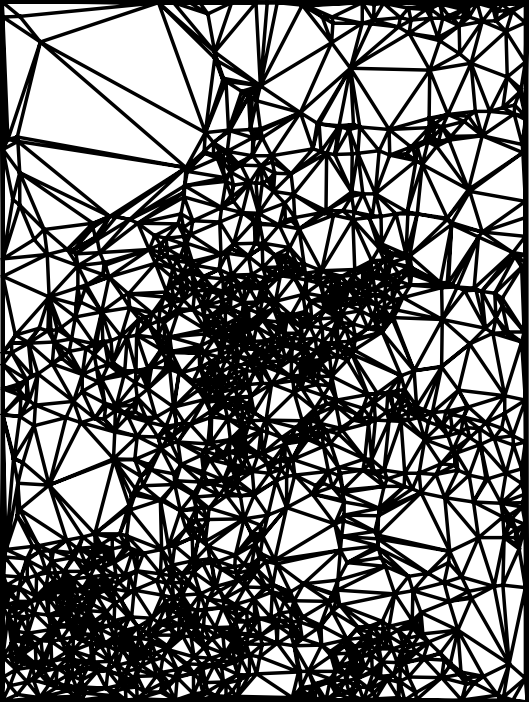}
        \includegraphics[width=\textwidth]{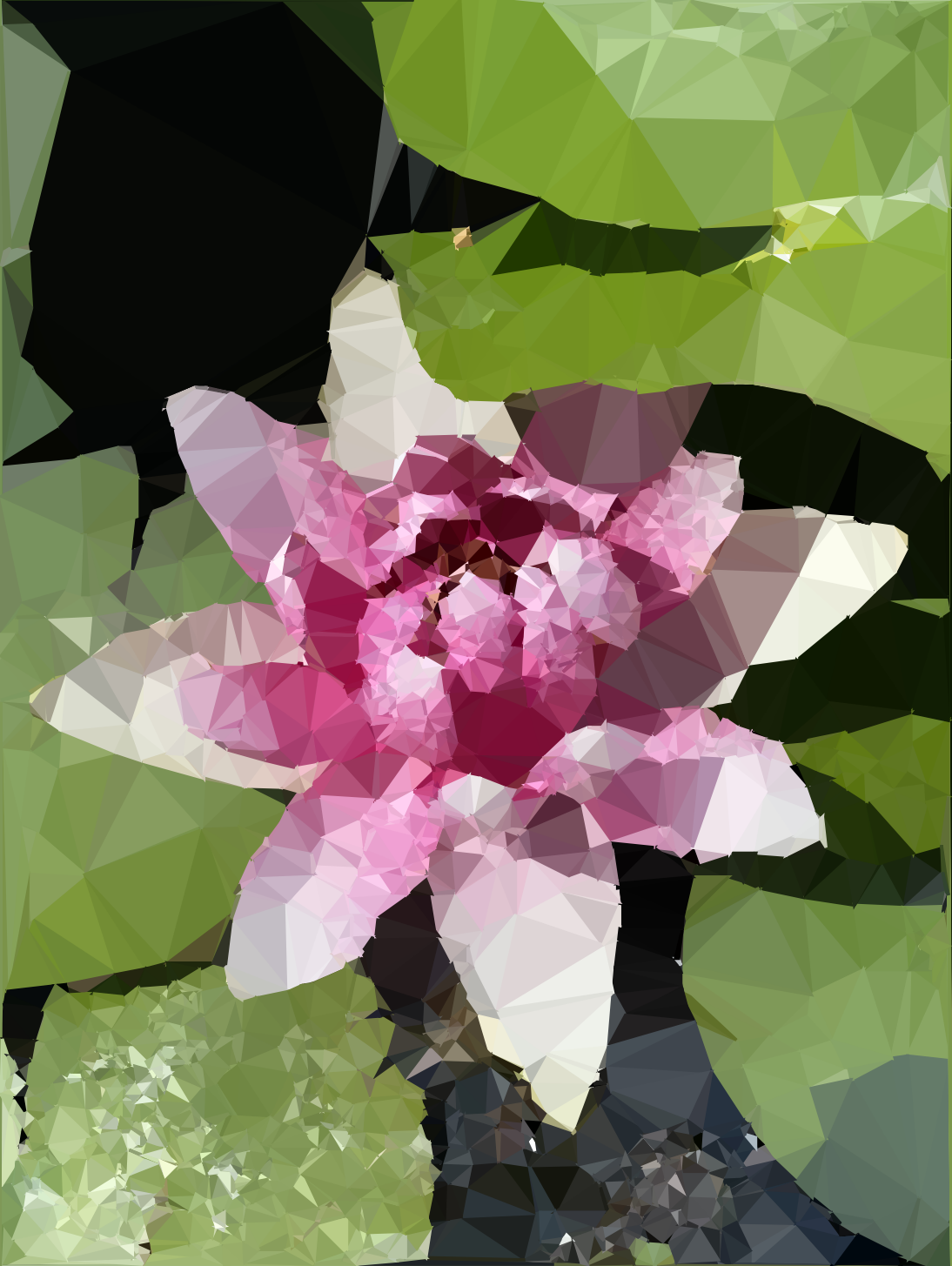}
        \subcaption{$t=25$}
    \end{subfigure}
    \begin{subfigure}[t]{0.3\textwidth}
        \centering
        \includegraphics[width=\textwidth]{readmeImages/waterLily_triangulation.png}
        \includegraphics[width=\textwidth]{readmeImages/waterLily_final.png}
        \subcaption{$t=50$}
    \end{subfigure}
        \begin{subfigure}[t]{0.3\textwidth}
        \centering
        \includegraphics[width=\textwidth]{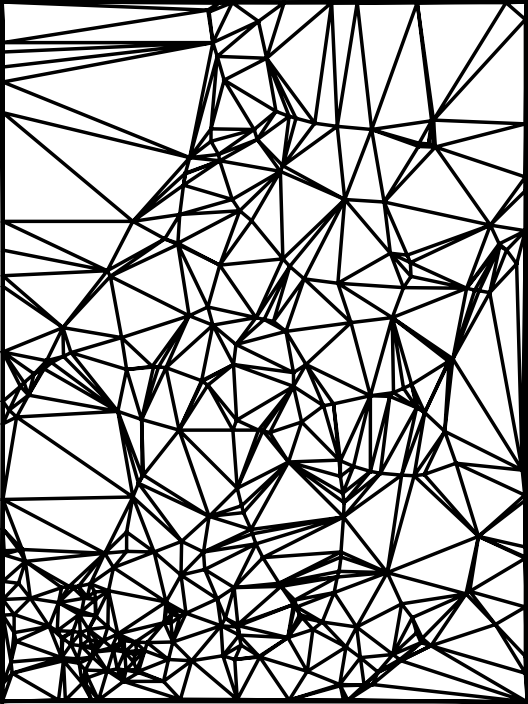}
        \includegraphics[width=\textwidth]{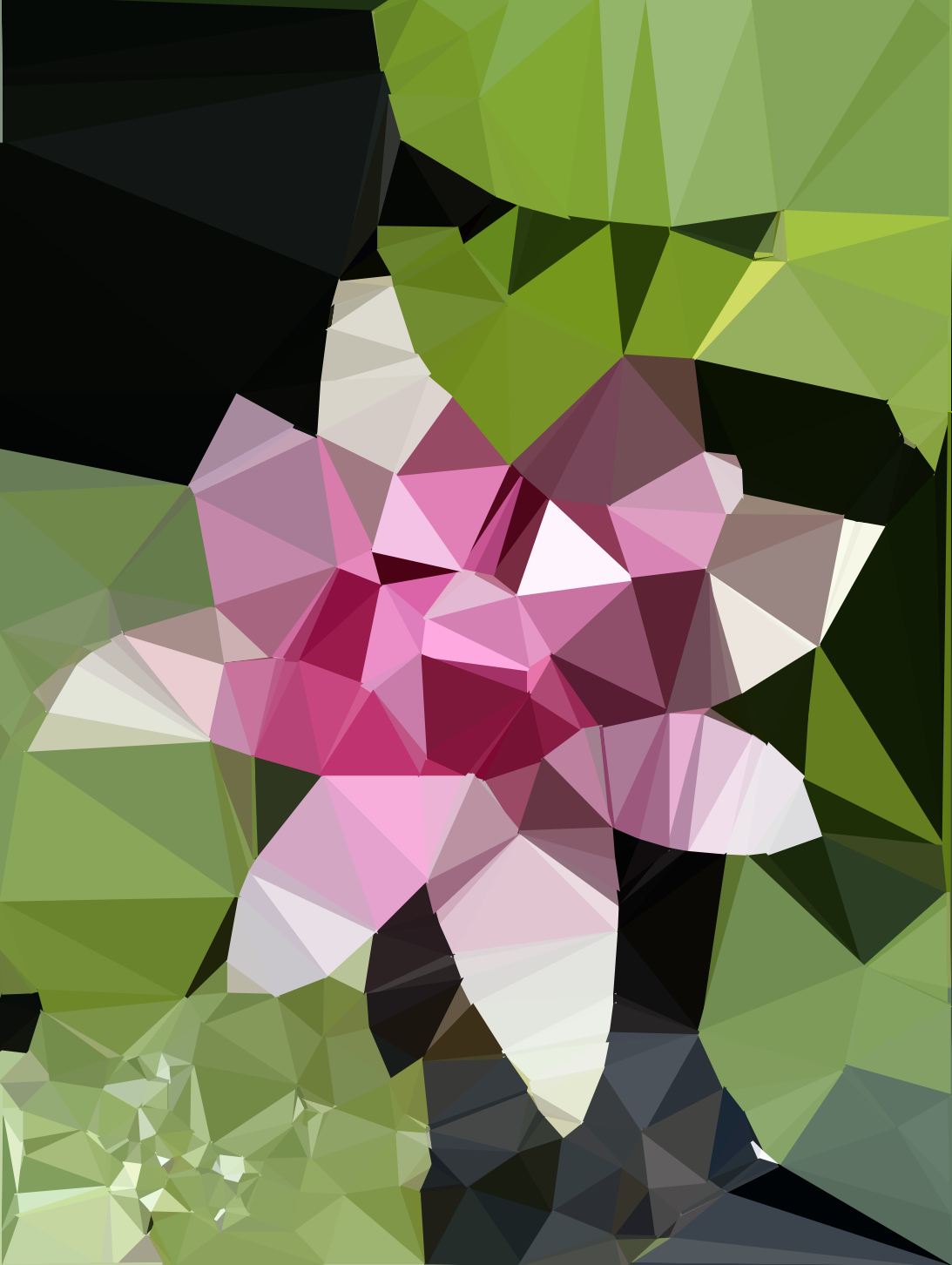}
        \subcaption{$t=75$}
    \end{subfigure}
    \caption{Differences in final image triangulation depending on threshold to select pixels.}
    \label{fig:threshold}
\end{figure}
Both are directly related to the number of vertices in the triangulation. As Figure \ref{fig:threshold} shows, the number of vertices and triangles decreases as the threshold increases for fixed $d$.
\begin{figure}[thbp]
    \begin{subfigure}[t]{0.3\textwidth}
        \centering
        \includegraphics[width=\textwidth]{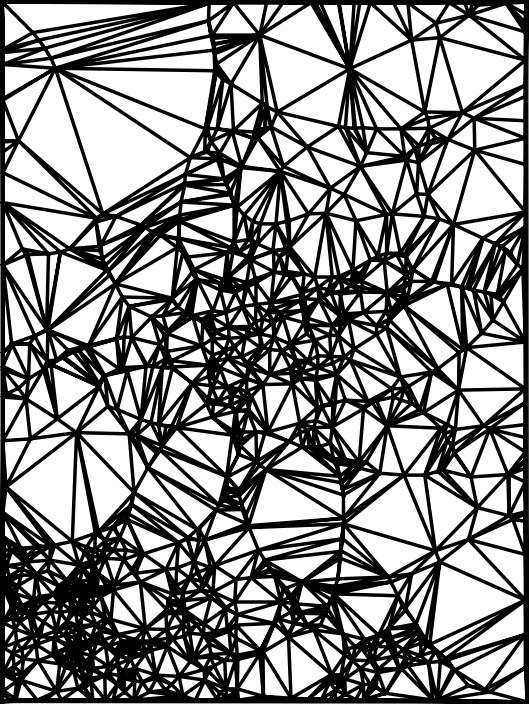}
        \includegraphics[width=\textwidth]{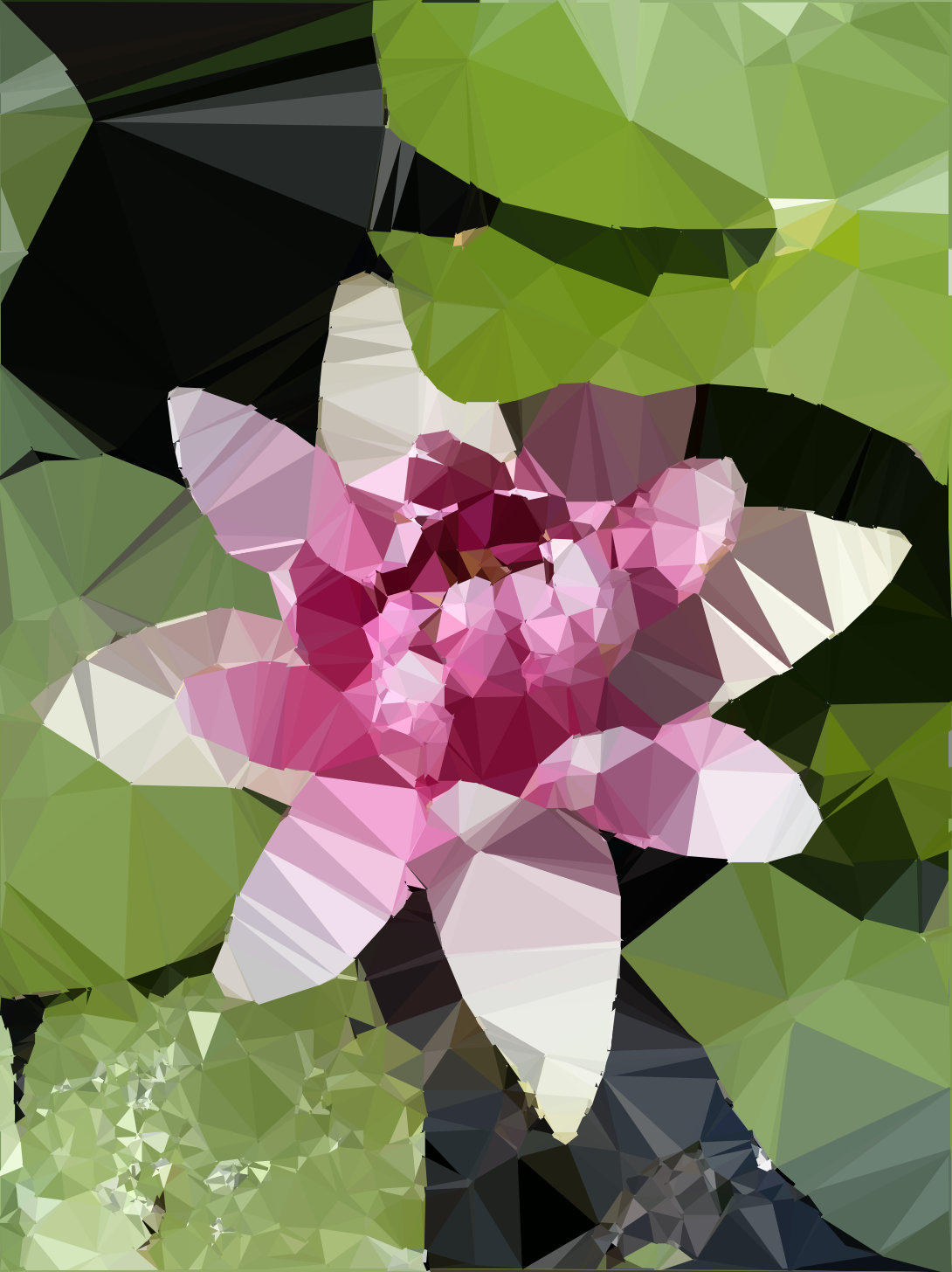}
        \subcaption{$d=35$}
    \end{subfigure}
    \begin{subfigure}[t]{0.3\textwidth}
        \centering
        \includegraphics[width=\textwidth]{readmeImages/waterLily_triangulation.png}
        \includegraphics[width=\textwidth]{readmeImages/waterLily_final.png}
        \subcaption{$d=60$}
    \end{subfigure}
        \begin{subfigure}[t]{0.3\textwidth}
        \centering
        \includegraphics[width=\textwidth]{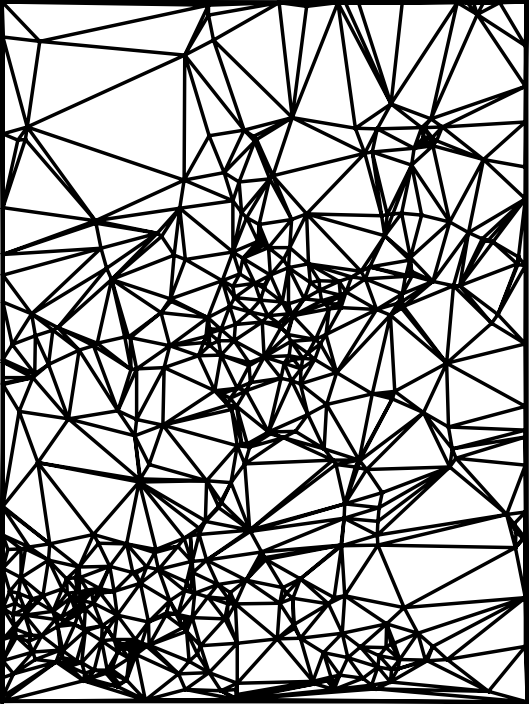}
        \includegraphics[width=\textwidth]{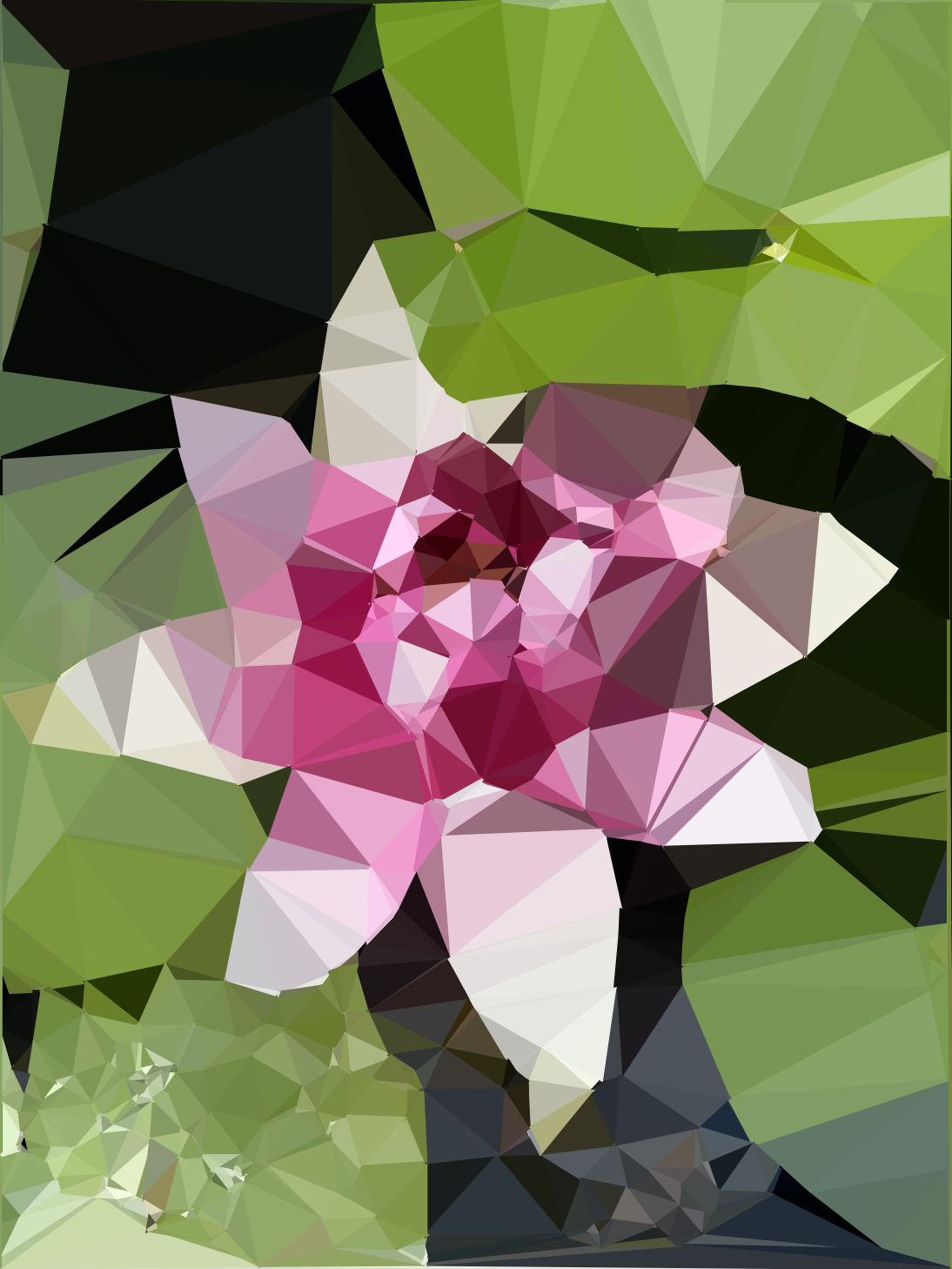}
        \subcaption{$d=85$}
    \end{subfigure}
    \caption{Differences in final image triangulation depending on density parameter to subsample pixels above the threshold.}
    \label{fig:density}
\end{figure}
Parameters $t$ and $d$ have similar effects. That is, increasing $d$ with fixed $t$ reduces the number of triangles and vertices, as seen in Figure \ref{fig:density}. The image in the leftmost panel in Figure \ref{fig:density}, though, demonstrates the extent to which $d$ declutters the points as opposed to $t$. As a result, the main purpose of density reduction is to decrease run time and avoid clusters of very small triangles.
\begin{figure}
    \centering
    \includegraphics[width=0.49\textwidth]{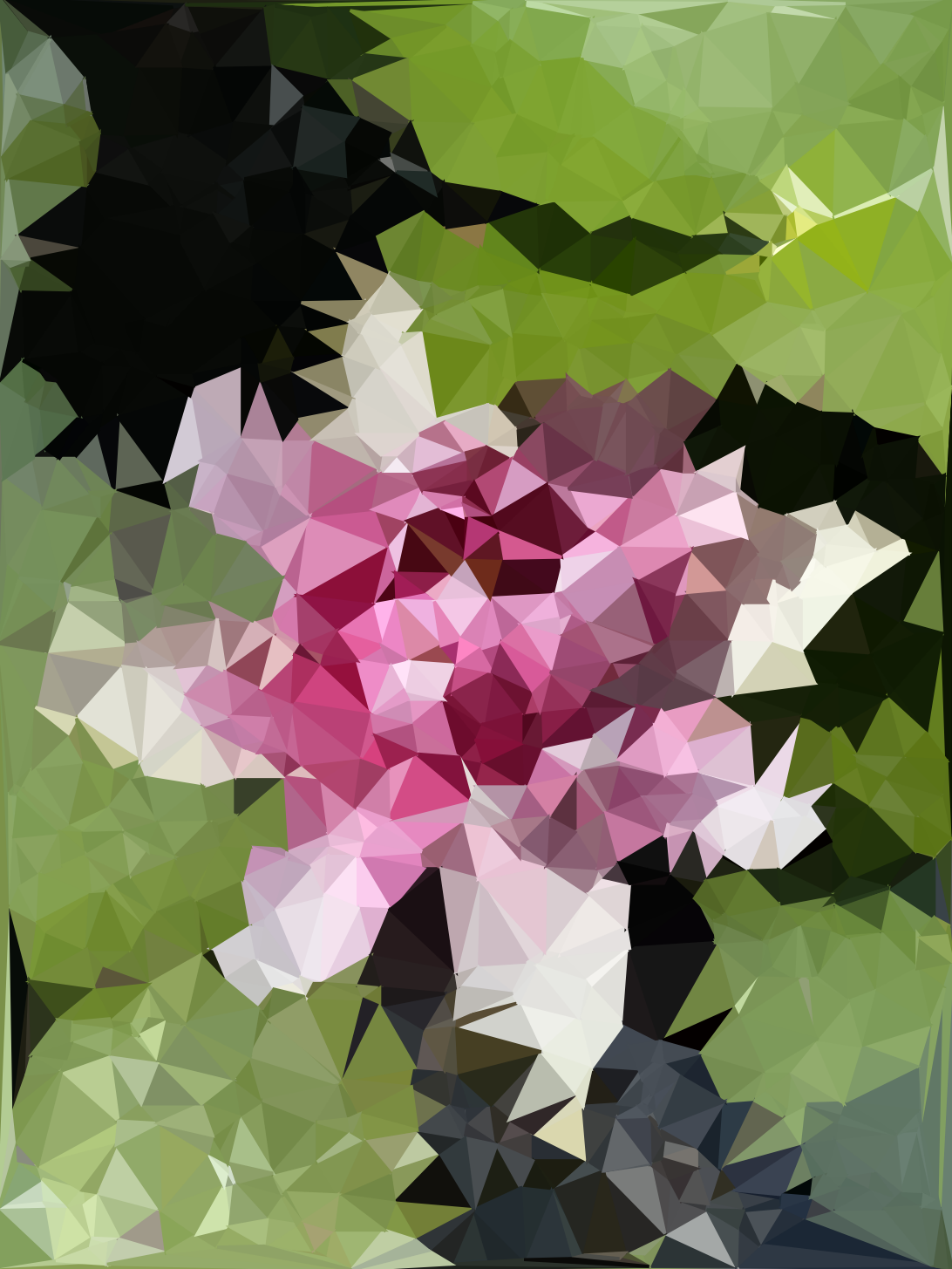}
    \includegraphics[width=0.49\textwidth]{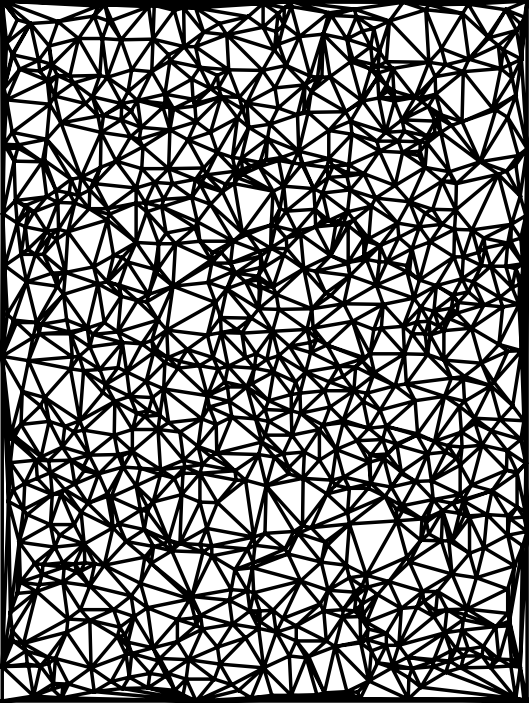}
    \caption{Triangulation using randomly selected pixels as vertices.}
    \label{fig:random}
\end{figure}

Finally, we consider random points as triangulation vertices as opposed to following the algorithm discussed here. We see in Figure \ref{fig:random} the image triangulation using 1000 randomly generated points, which is slightly more than the number of vertices that our algorithm finds (with $t=50$ and $d=60$). The integrity of the image is destroyed, losing many of the features. By using edge detection, though, we can reduce the number of points needed to make a recognizable image and preserve the underlying skeleton.

\section{Conclusion}
In this brief article, we describe an algorithm to triangulate an image, detailed in the Github repository \cite{github}. This repository also contains a video of several examples of resulting artistic triangulated images. While the algorithm we outline successfully triangulates any image, the ideal threshold value and density reduction parameter are subjective. If a user desires a more abstract image, a higher threshold value, higher density reduction parameter, or randomized point cloud is suitable. However, if a user is aspiring for a triangulated image closer to the original image, the opposite holds true. Alternatively to an image triangulation, a user may instead wish to consider the dual graph of the Delaunay triangulation. Each Voronoi region would then be colored accordingly instead of each triangle. The output would be more similar to a mosaic and could be considered a separate form of art.

\section{Acknowledgements}
Support through NSF:CDS\&E-MSS-1854703 and NSF:BCS-2318171.

\bibliographystyle{plainurl}
\bibliography{arxiv}

\end{document}